# APPLICATION OF MULTIPHASE PARTICLE METHODS IN ATOMIZATION AND BREAKUP REGIMES OF LIQUID JETS

**Amirsaman Farrokhpanah, Javad Mostaghimi**
Department of Mechanical & Industrial Engineering, University of Toronto
Toronto, Ontario, Canada

## ABSTRACT
Multiphase Smoothed Particle Hydrodynamics (SPH) method has been used to study the jet breakup phenomena. It has been shown that this method is well capable of capturing different jet breakup characteristics. The value obtained for critical Weber number here in transition from dripping to jetting is a very good match to available values in literature. Jet breakup lengths are also agreeing well with several empirical correlations. Successful usage of SPH, as a comparably fast CFD solver, in jet breakup analysis helps in speeding up the numerical study of this phenomenon.

## INTRODUCTION
Spray and atomization systems are inseparable part of many engineering applications from inkjet printers, respiratory drug administration, molecular biology analysis, aerosol fuel delivery in combustion engines and sample introduction in atomic and mass spectrometry. Numerical simulation of these phenomena might be cumbersome due to the amount of computations involved. Many of the methods available are mesh-dependent and might demand low resolutions to generate accurate results. Smoothed Particle Hydrodynamics method (SPH), is a Lagrangian mesh-free CFD method that can be advantageous in improving the runtime of these types of simulations. SPH, introduced in 1977 by Lucy [1] and Gingold and Monaghan [2], was initially used in astrophysics studies. Subsequently many studies have successfully applied SPH on fluid simulations like the work of Morris et al. [3] in 1997. Since the pressure Poisson equation is not solved in most SPH algorithms, the amount of runtime is significantly decreased. This makes SPH an influential tool capable of generating real-time results specially when used in single phase form on free surface fluids [4]. In 2006, Hu et al. [5] proposed a reliable approach for multiphase SPH that has been used ever since in many studies. Although this method is accurate in multiphase simulations where surface tension forces are dominant, it requires small time steps for large density and viscosity ratios between fluid phases.

In this study, a multiphase SPH solver based on the mentioned method has been used. To make this solver efficient, it has been optimized for execution on Graphic processing Units (GPUs). Results show that the jet breakup and atomization can be robustly simulated using this method while runtime can be significantly lower than other common CFD methods. Single phase SPH methods have been previously used to study the jet breakup phenomena like the works of Sirotkin et al. [6] and Takashima et al. [7]. To the best of authors' knowledge, no study has focused on investigating the ability of multiphase SPH in capturing characteristics of the jet breakup phenomena.

## GOVERNING EQUATIONS
Incompressible and isothermal Navier-Stokes equations in Lagrangian form are

$$\frac{d\rho}{dt} = -\rho \boldsymbol{\nabla} \cdot \boldsymbol{V} \qquad (1)$$

$$\frac{d\boldsymbol{V}}{dt} = \frac{1}{\rho}[-\nabla p + \mu \nabla^2 \boldsymbol{V} + \boldsymbol{F}^{st} + \boldsymbol{F}^b] \qquad (2)$$

with $\boldsymbol{F}^b$ being external body force like gravity. The surface tension force, $\boldsymbol{F}^{st}$, is approximated based on the Continuum Surface Force (CSF) model of Brackbill et al. [8], and for the case of constant surface tension is given by

$$\boldsymbol{F}^{st} = -\alpha \kappa \delta_s \hat{\boldsymbol{n}} \qquad (3)$$

here, $\alpha$ is the surface tension coefficient, $\kappa$ the local interface curvature, and $\hat{\boldsymbol{n}}$ the surface normal. Instead of solving the Poisson equation for pressure, equation of state suggested by Batchelor [9] is used which calculates pressure using density variations in the form of

$$P = P_0 \left(\frac{\rho}{\rho_0}\right)^\gamma + b \qquad (4)$$

$\gamma$, resembling the gas constant is typically taken to be 7 for liquids and 1.4 for vapor phases. $b$ is a constant background pressure, and $p_0$ represents a reference pressure chosen to keep maximum density variations from $\rho_0$ in the order of O (1%). This low variation of density assures that the system remains almost incompressible while simultaneously possessing compressible features which allow the usage of equation of state.

## SMOOTHED PARTICLE HYDRODYNAMICS MULTI-PHASE MODEL

Equations described above are discretized using particles in SPH. These particles are moved at each time step in a Lagrangian manner based on their new calculated velocities. Integral interpolations are used to calculate different variables at a particle position from its neighboring particles. More detailed discussions on SPH formulation and derivations can be found at [10].

The density of particle $i$, can be evaluated from neighboring particles, $j$ from [11]

$$\rho_i = m_i \sum_j W_{ij} = m_i \, \sigma_i \qquad (5)$$

where $m_i$ is the mass of each SPH particle and $W_{ij}$ a quintic spline kernel [3] between particle $i$ and its neighboring $j$ particles defined as [12]

$$W_{ij} = \alpha^* \begin{cases} (3-R)^5 - 6(2-R)^5 + 15(1-R)^5 & 0 \le R < 1 \\ (3-R)^5 - 6(2-R)^5 & 1 \le R < 2 \\ (3-R)^5 & 2 \le R < 3 \\ 0 & 3 \le R \end{cases} \qquad (6)$$

where $R = r_{ij}/h$, with $r_{ij}$ being the distance between the two particles $i$ and $j$. $h$ is the smoothing length or cutoff distance which determines how large the neighborhood of a particle is. $\alpha^*$ can be calculated from normalization of the kernel $W_{ij}$ and here is $120/h$, $7/478\pi h^2$, and $3/359\pi h^3$ for one, two, and three dimensional spaces, respectively. The advantage of using equation (5) over other common methods of calculating density in SPH is that since the density of particle $i$ only depends on the mass of this particle ($m_i$), density is not smoothed near the regions were two phases in a multiphase environment meet, and therefore, equation (5) has the ability of reproducing sharp density jump between phases. In this way, each particle treats all neighbors as if they have the same rest density and mass as itself [13].

As Hu et al. [5] proposed, spatial derivatives for multiphase SPH simulations for a smoothed variable like $\psi$ can be in the form of

$$\nabla \psi_i = \sum_j \left(\frac{\psi_i}{\sigma_i^2} + \frac{\psi_j}{\sigma_j^2}\right) \sigma_i \frac{\partial W_{ij}}{\partial r_{ij}} \boldsymbol{e}_{ij} \qquad (7)$$

where $\boldsymbol{e}_{ij}$ is the normal vector passing from particle $i$ to $j$. Applying equation (7) to pressure derivative term in equation (2), the pressure acceleration of the momentum equation for particle $i$ is [5]

$$\boldsymbol{F}_i^p = -\frac{1}{m_i} \sum_j \left(\frac{p_i}{\sigma_i^2} + \frac{p_j}{\sigma_j^2}\right) \frac{\partial W_{ij}}{\partial r_{ij}} \boldsymbol{e}_{ij} \qquad (8)$$

Considering two fluid phases like $k$ and $l$, the averaged shear stress between particles can be calculated to be

$$\overline{\Pi_{ij}^v} = \frac{2\mu^k \mu^l}{r_{ij}(\mu^l + \mu^k)} \left(\boldsymbol{e}_{ij} \boldsymbol{V}_{ij} + \boldsymbol{V}_{ij} \boldsymbol{e}_{ij}\right) \qquad (9)$$

Combining equations (7) and (9), the viscous force in equation (2) can be rearranged in the form of

$$\boldsymbol{F}^v = \frac{1}{m_i} \sum_j \frac{2\mu^k \mu^l}{(\mu^l + \mu^k)} \left(\frac{1}{\sigma_i^2} + \frac{1}{\sigma_j^2}\right) \frac{\boldsymbol{V}_{ij}}{r_{ij}} \frac{\partial W_{ij}}{\partial r_{ij}} \qquad (10)$$

To calculate the surface force, a color function $C_i^s$ is defined with a value of 1 when a particle is inside arbitrary phase $s$ and otherwise zero. Using equation (7), the gradient of this function for particle $i$ of phase $k$ is calculated as

$$\nabla C_i^{k,l} = \sigma_i \sum_j \left(\frac{C_i^l}{\sigma_i^2} + \frac{C_j^l}{\sigma_j^2}\right) \frac{\partial W_{ij}}{\partial r_{ij}} \boldsymbol{e}_{ij} \qquad (11)$$

By taking equation (3) in the tensor form of $\boldsymbol{F}^{st} = \boldsymbol{\nabla} \cdot \Pi$, the surface stress tensor $\Pi$ for the two phases of $k$ and $l$, becomes [5]

$$\Pi = \alpha \left(\frac{1}{d} \boldsymbol{I} - \hat{n}\hat{n}\right) |\nabla C| \qquad (12)$$

where $d$ is the is 1, 2, 3 for one, two and three dimensions, respectively and $\boldsymbol{I}$ is the unit tensor. By defining $\hat{\boldsymbol{n}} = \nabla C / |\nabla C|$, equation (12) beomces

$$\Pi_i^{k,l} = \alpha^{k,l} \frac{1}{|\nabla C_i^{k,l}|} \left(\frac{1}{d} I |\nabla C_i^{k,l}|^2 - \nabla C_i^{k,l} \nabla C_i^{k,l}\right) \quad (13)$$

By summing all tensors between particle $i$ and other phases, the total surface stress tensor is gained as

$$\Pi_{i\in k}^{total} = \sum_{\forall l \neq k} \Pi_i^{k,l} \quad (14)$$

By utilizing equations (7) and (14), equation (3) for particle $i$ is transformed into

$$F^{st} = \frac{1}{m_i} \sum_j \frac{\partial W_{ij}}{\partial r_{ij}} e_{ij} \left(\frac{\Pi_i^{total}}{\sigma_i^2} + \frac{\Pi_j^{total}}{\sigma_j^2}\right) \quad (15)$$

## RESULTS

Here, jet breakup of a liquid exiting a nozzle is investigated. Figure 1, shows particle positioning during the simulation. Initially particles are position at the top left inside the nozzle. Black particles in this figure show the wall of the nozzle. The y-axis is used as the symmetry line. A parabolic velocity field is constantly applied to the fluid inside the nozzle. After exiting the nozzle, gravitational force of 9.81m/s² is responsible for accelerating the fluid. Full domain height has not been shown in this figure. It has been chosen to be approximately twice the size of the breakup length. The width of the domain is chosen to be four times of the nozzle radius.

For the sake of comparison, test case specifications are chosen similar to those of Sirotkin et al. [6]. The fluid is chosen to be Diethylene glycol (DEG) with $\rho = 1120$ kg/m³, $\mu = 3.16E-2$ kg m⁻¹ s⁻¹, $\alpha = 44E-3$ N/m (all at 23°C). Nozzle diameter is $D = 5E-3$ m. The following non-dimensional numbers are used for better understanding of the results: Weber number ($We = \rho \bar{v}^2 R/\alpha$), Reynolds number ($Re = \rho \bar{v} R/\mu$), and Froude number ($Fr = v/\sqrt{gD}$), with $R$ being the nozzle radius. Average inlet velocity of fluid at nozzle exit ($\bar{v}$) is changed while the nozzle diameter and gravity forces remain the same creating test cases shown in table 1.

**Table 1. Test case specifications**

| Case # | $\bar{v}$ (m/s) | Re | We | Fr |
|---|---|---|---|---|
| 1 | 3.97E−2 | 3.52 | 0.100 | 0.18 |
| 2 | 5.62E−2 | 4.98 | 0.201 | 0.25 |
| 3 | 6.88E−2 | 6.07 | 0.301 | 0.31 |
| 4 | 7.94E−2 | 7.03 | 0.401 | 0.36 |
| 5 | 9.73E−2 | 8.62 | 0.60 | 0.44 |
| 6 | 12.56E−2 | 11.13 | 1.00 | 0.57 |
| 7 | 17.76E−2 | 15.74 | 2.01 | 0.80 |
| 8 | 21.75E−2 | 19.27 | 3.01 | 0.98 |
| 9 | 25.12E−2 | 22.26 | 4.02 | 1.13 |

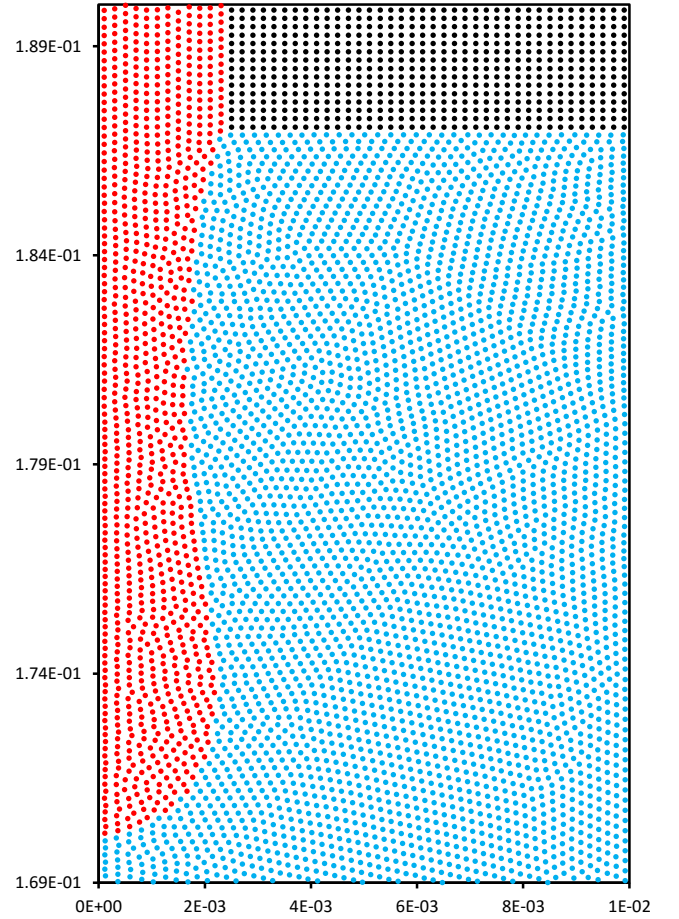

**Figure 1.** Particle positions during the solution: fluid particles colored in red exiting the nozzle colored in black. Air particles surrounding the fluid are colored blue.

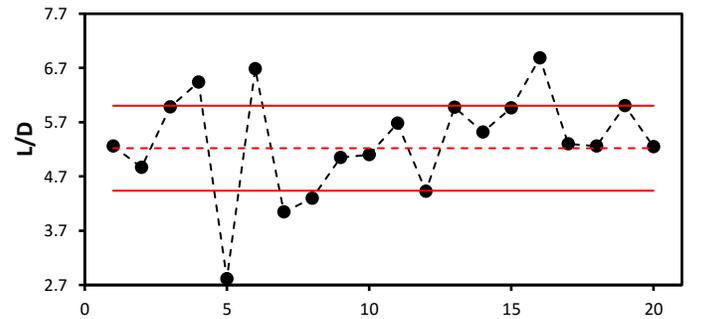

**Figure 2.** Breakup length over nozzle diameter for We=0.1, Re=3.52

Figure 2,3, and 4 show the non-dimensional breakup length (L/D) for cases #1, #2 , and #3, respectively. L, is the length of the liquid jet outside the nozzle right before it breaks (can also be shown as $L_b$, the breakup length). D is the nozzle diameter. The x-axis in these figures shows the counter for the successive breakups as time proceeds.

The dashed red line in these figure, shows the following correlation obtained from analytical analysis of viscous jet breakup [14] for a jet with no shear stress (exiting in vacuum or a low-speed jet exiting in a gas) [15]

$$\frac{L}{D} = C(We^{1/2} + \frac{3We}{Re}) \qquad (16)$$

Experimental results suggest that for the case of low viscosity jets, $C$ can have a value of 13 [14]. Other reported values for C are 12, 13, and 13.4 [15]. Solid red lines in these figures indicate ±15% deviation from the red dashed line.

These test cases show that for lower fluid exit velocities, shown in figures 2 and 3, where Weber number is 0.1 and 0.2, the breakup lengths are very close to the predictions made by equation 16. These Weber number values correspond to a liquid jet behavior know as dripping. In this region, gravitational forces are more dominant compared to inertia forces acting on the jet. A critical Weber number between 0.28-0.33 has been reported in literature [6]. Above the critical Weber number, jet behavior changes from dripping to jetting. Results show that for higher velocities such as the one shown in figure 4 related to We=0.3, deviation from equation 16 becomes significant. That might be due to the face that equation 16 has been derived with assumption of liquid jets being at low speeds and shear stresses working on the jet surface should not be significant.

Another correlation relating the non-dimensional number is in the form of [14]

$$\frac{L}{D} = 1.04C\sqrt{We} \qquad (17)$$

C has been chosen again to be 13 based on experimental results [14]. Test cases here demonstrate that breakup lengths for larger velocities are closer to values predicted by equation 17 rather than equation 16. Results for several velocities have been shown in figure 5. Red and blue lines in this figure show predictions of equation 16 and 17, respectively. Black dots show multiphase SPH results obtained here.

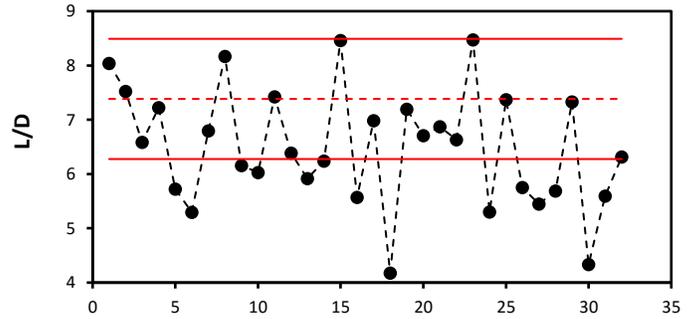

Figure 3. Breakup length over nozzle diameter for We=0.2, Re=4.98

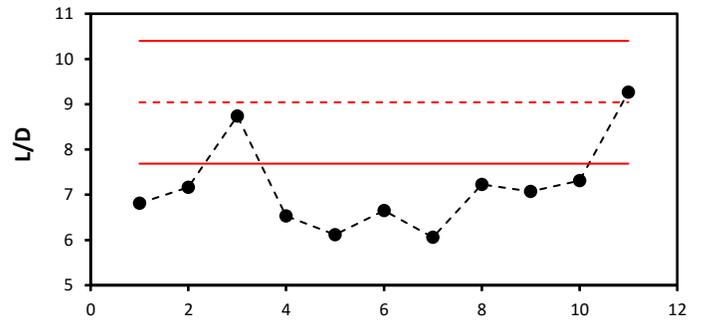

Figure 4. Breakup length over nozzle diameter for We=0.3, Re=6.07

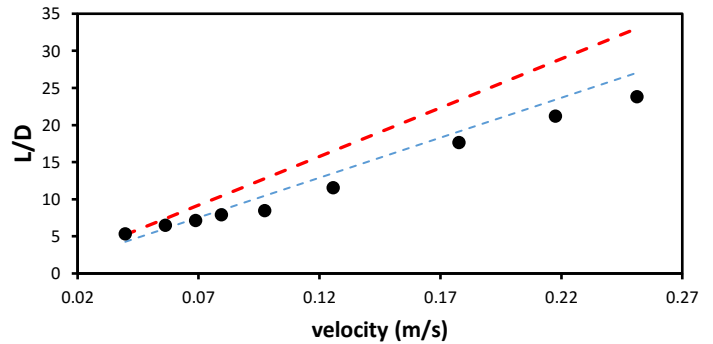

Figure 5. Comparison of breakup lengths for various velocities on a constant nozzle diameter. Multiphase SPH results (black dots) versus equation 16 (red line) and equation 17 (blue line)

Drop sizes for the cases discussed have been shown in figures 6-8. It can be confirmed that overall consistent drop sizes are obtained in the simulation here. Major drop sizes for most of the cases of small Weber numbers remains close to each other and the nozzle diameter. Satellite drops, shown by red dots in figures, possess radius of almost half of the major drops.

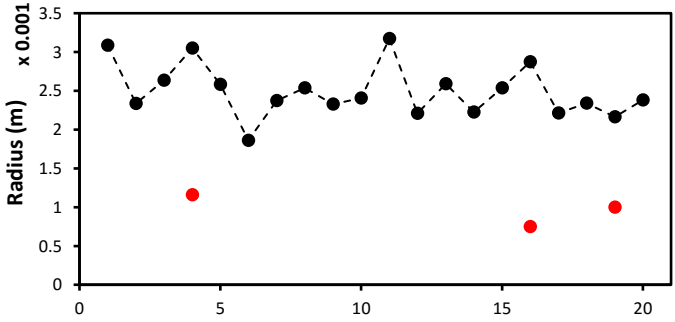

Figure 6. Radius of generated drops (black dots) and satellite drops (red dots) for We=0.1, Re=3.51

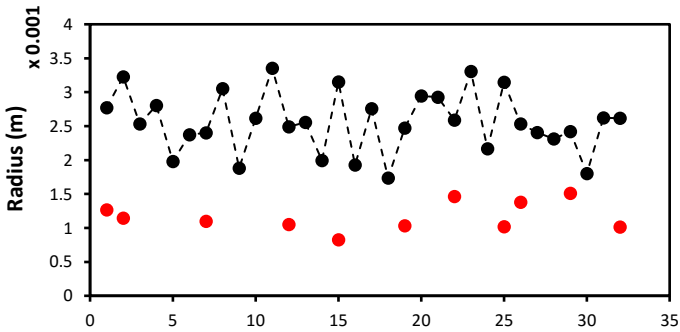

Figure 7. Radius of generated drops (black dots) and satellite drops (red dots) for We=0.2, Re=4.98

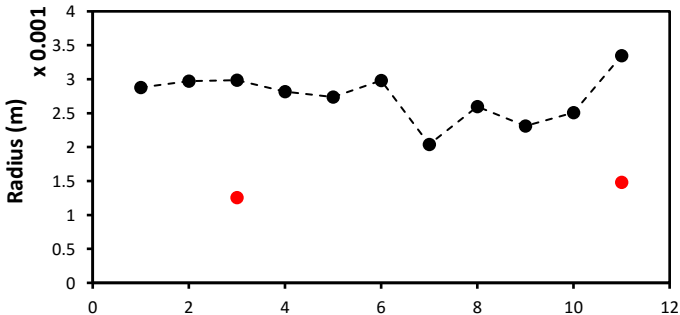

Figure 8. Radius of generated drops (black dots) and satellite drops (red dots) for We=0.3, Re=6.07

Time between jet breakups has been plotted in figures 9-11. It can be seen that for each test case, breakups occur in approximately the same time intervals. It is also noticeable that breakup interval times for Weber numbers below the critical values are almost identical while when We is increased above the critical value in the case of We=0.3, time intervals are significantly larger (almost ten times). This clearly indicates transition from dripping to jetting where the gravity force is no longer the dominant breakup factor as inertia becomes larger.

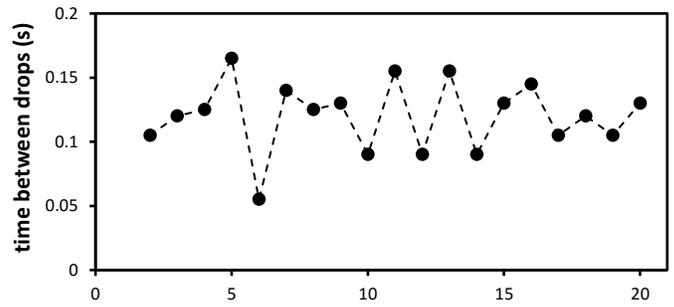

Figure 9. Time interval between jet breakups for We=0.1, Re=3.51

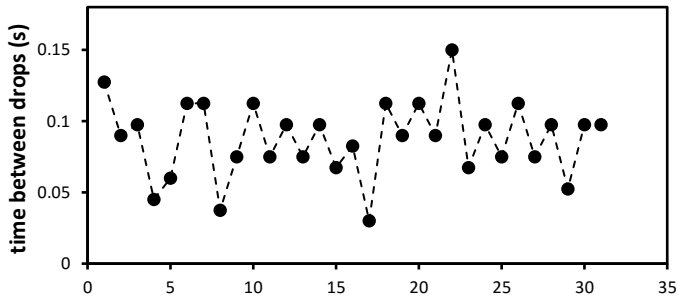

Figure 10. Time interval between jet breakups for We=0.2, Re=4.98

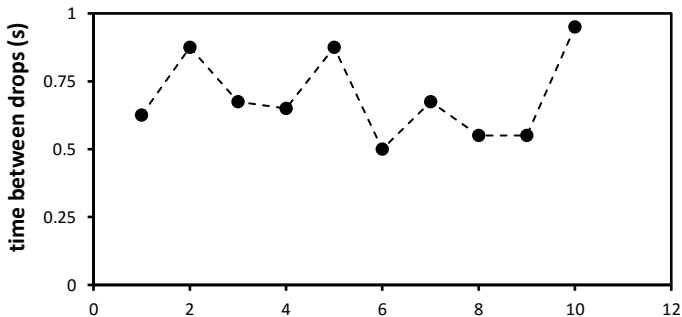

Figure 11. Time interval between jet breakups for We=0.3, Re=6.07

## CONCLUDING REMARKS

Test cases reported here validate the usage of multiphase SPH solver in capturing the jet breakup phenomena. Transition from dripping to jetting was apparent to be between Weber numbers of 0.2 and 0.3 which matches the reported values in literature. Dripping jet results matched well with the experimental correlation (equation 16) derived for low-velocity viscous jets.

For larger velocities, results start to deviate from this correlation. It has been found out that by assigning a value of 10 for the empirical variable C in equation 16, results obtained here match well with equation 16 for all ranges of Weber number for both dripping and jetting regimes. Equation 17 seems to be better in predicting the results obtained numerically here.

**REFERENCES**

1. *A numerical approach to the testing of the fission hypothesis.* **Lucy, L.B.** 1977, Astron. J., Vol. 82, pp. 1013–1024.

2. *Smoothed particle hydrodynamics.* **Gingold, R. A. and Monaghan, J. J.** 1977, pp. 375-389.

3. *Modeling low Reynolds number incompressible flows using SPH.* **Morris, J.P., Fox, P.J. and Zhu, Y.** 1, School of Civil Engineering, Purdue University, West Lafayette, Indiana : Journal of Computational Physics, 1997, Vol. 136.

4. *Simulating free surface flows with SPH.* **Monaghan, J. J.** 2, Feb. 1994, Journal of Computational Physics, Vol. 110, pp. 399-406.

5. *A multi-phase SPH method for macroscopic and mesoscopic flows.* **Hu, X.Y. and Adams, N.A.** 213, s.l. : Elsevier Inc., 2006, Journal of Computational Physics, pp. 844–861.

6. *A new particle method for simulating breakup of liquid jets.* **Sirotkin, Fedir V. and Yoh, Jack J.**

7. *Simulation of Liquid Jet Breakup Process by Three-Dimensional Incompressible SPH Method.* **Takashima, T., et al., et al.** Big Island, Hawaii : s.n., 2012. Seventh International Conference on Computational Fluid Dynamics (ICCFD7).

8. *A continuum method for modeling surface tension.* **Brackbill, J.U, Kothe, D.B and Zemach, C.** 1992, Journal of Computational Physics, pp. 335-354.

9. **Batchelor, G. K.** *An introduction to fluid dynamics.* s.l. : Cambridge University Press, 1967.

10. **Farrokhpanah, A.** *Applying Contact Angle to a Two-Dimensional Smoothed Particle Hydrodynamics (SPH) model on a Graphics Processing Unit (GPU) Platform.* s.l. : University of Toronto, 2012.

11. *Smoothed particle hydrodynamics.* **Monaghan, J.** s.l. : INSTITUTE OF PHYSICS PUBLISHING, 2005.

12. **Liu, G. R. and Liu, M. B.** *Smoothed Particle Hydrodynamics: A mesh free particle method.* Singapore : World Scientific, 2003.

13. *Density Contrast SPH Interfaces.* **Solenthaler, B. and Pajarola, R.** Visualization and MultiMedia Lab, University of Zurich, Switzerland : Eurographics/ ACM SIGGRAPH Symposium on Computer Animation, 2008.

14. **Ashgriz, Nasser, [ed.].** *Handbook of Atomization and Sprays.* New York : Springer, 2011.

15. *Newtonian jet stability.* **Grant, Rollin Peter and Middleman, Stanley.** 4, 1966, AIChE Journal, Vol. 12, pp. 669–678.